\documentclass[12pt]{article}
\usepackage{amssymb,amsmath,epsfig}

\begin{document}

\title{\bf Energy Contents of Some Non-Vacuum Spacetimes in Teleparallel
Gravity}
\author{M. Sharif\thanks {msharif@math.pu.edu.pk} and Sumaira
Taj\thanks{sumairataj@ymail.com}\\
Department of Mathematics, University of the Punjab,\\
Quaid-e-Azam Campus, Lahore-54590, Pakistan.}

\date{}

\maketitle
\begin{abstract}
This paper elaborates the problem of energy-momentum in the
framework of teleparallel equivalent of General Relativity. For this
purpose, we consider energy-momentum prescription derived from the
integral form of the constraint equations developed in the
Hamiltonian formulation of the teleparallel equivalent of General
Relativity. We use this technique to investigate energy-momentum of
stationary axisymmetric Einstein-Maxwell solutions and cosmic string
spacetimes. The angular momentum, gravitational and matter
energy-momentum fluxes of these spacetimes are also evaluated. It is
concluded that the results of teleparallel theory are relatively
analogous to the results of General Relativity.
\end{abstract}


\section{Introduction}

The localization problem of energy-momentum emerged along with the
field equations and still inconclusive. The conservation laws of
energy and momentum for gravitation are major causes of the problem.
The localization is not debatable for flat spacetime. However, for
curved spacetime, the energy-momentum tensor of matter plus
gravitation fields no longer satisfies the conservation law. Various
energy-momentum prescriptions have been proposed to overcome this
problem. The first of such contributions came from Einstein
\cite{1}, who gave an expression for energy-momentum due to matter,
non-gravitational and gravitational fields. After this, M{\o}ller
\cite{2}, Landau-Lifshitz \cite{3}, Papapetrou \cite{4}, Bergmann
\cite{5}, Tolman \cite{6}, Weinberg \cite{7} and Komar \cite{8}
proposed their prescriptions. The key issue with energy-momentum
prescriptions is their coordinate dependence and non-uniqueness. All
these prescriptions, except M{\o}ller and Komar, give acceptable
results only when one uses Cartesian coordinates.

Virbhadra and his collaborators \cite{12}-\cite{13} considered
several spacetimes and verified that different prescriptions could
give the same results for a given spacetime. Radinschi \cite{14}
evaluated energy distribution of Bianchi type I universe and found
the same results as those of Banerjee and Sen \cite{15} and Xulu
\cite{16}. In a recent paper, Abbassi et al. \cite{17} have proved
that various prescriptions provide the same results in static and
non-static cosmic string spacetimes. Sharif \cite{18} have shown
that there exist several spacetimes that do not yield the same
results for different prescriptions.

Some authors \cite{21}-\cite{22} argued that the localization
problem of energy-momentum becomes more transparent in the framework
of teleparallel equivalent of General Relativity (GR). M{\o}ller
\cite{23} was the first who observed that the tetrad description of
the gravitational field could lead to a better expression for the
gravitational energy-momentum than does GR. Mikhail et al. \cite{24}
derived the energy-momentum complex in M{\o}ller's tetrad theory.
Loi So and Vargas \cite{25} investigated that the quasi-local
energy-momentum of Bianchi types I and II universes vanishes
everywhere in the context of teleparallel gravity which coincides
with the result in GR. In recent papers, Sharif and Amir
\cite{26}-\cite{27} concluded that in teleparallel theory, the
results of energy-momentum prescriptions did not coincide with GR
for a given spacetime.

Andrade et al. \cite{28} considered the localization of energy in
Lagrangian framework of teleparallel equivalent GR (TEGR). Maluf et
al. \cite{29} derived the expression for the gravitational energy,
momentum and angular momentum from the Hamiltonian formulation of
TEGR \cite{30}. Maluf computed gravitational energy of the de Sitter
spacetime \cite{31} and gravitational pressure for the Schwarzschild
spacetime \cite{32} in the framework of TEGR. da Rocha-Neto and
Castello-Branco \cite{33} extended this procedure to evaluate
gravitational energy of the Kerr and Kerr anti-de Sitter spacetimes.
In a recent paper, Maluf and Ulhoa \cite{34} and the authors
\cite{34a} discussed gravitational energy-momentum of gravitational
waves.

In this paper, we elaborate the above procedure by evaluating energy
and its relevant quantities for stationary axisymmetric solutions of
the Einstein-Maxwell field equations and cosmic string spacetimes.
The paper has been organized in the following fashion: Section $2$
presents some elementary notions of TEGR and energy-momentum
expressions. In section $3$, we calculate energy and its contents
for stationary axisymmetric solutions. Section $4$ is devoted for
the two special cases of  stationary axisymmetric solutions of the
Einstein-Maxwell field equations. Energy and its contents for
cosmic string spacetimes are discussed in section $5$. In
section $6$, we summarize and discuss the results.

Here, the spacetime indices $(\mu,\nu,\rho,...)$ and tangent space
indices $(a,b,c,...)$ run from 0 to 3. Time and space indices are
denoted according to $\mu=0,i$, $a=(0),(i)$.

\section{Hamiltonian Approach: Energy-Momentum in
Teleparallel Theory}

The torsion tensor in terms of tetrad field is defined as
\begin{equation}\label{2.3}
{T^a}_{\mu\nu}=\partial_\mu{e^a}_\nu-\partial_\nu{e^a}_\mu,
\end{equation}
which is related to the Weitzenb\"{o}ck connection \cite{35}
\begin{equation}\label{2.4}
{\Gamma^\lambda}_{\mu\nu}={e_a}^\lambda\partial_\nu{e^a}_\mu.
\end{equation}
The Lagrangian density for the gravitational field in TEGR is
\cite{30}
\begin{eqnarray}\label{2.5}
L=-\kappa
e(\frac{1}{4}T^{abc}T_{abc}+\frac{1}{2}T^{abc}T_{bac}-T^{a}T_{a})-L_M
\equiv-\kappa e\Sigma^{abc}T_{abc}-L_M,
\end{eqnarray}
where $\kappa=1/16\pi$ and $e=det({e^a}_\mu)$. The tensor
$\Sigma^{abc}$ is defined as
\begin{equation}\label{2.6}
\Sigma^{abc}=\frac{1}{4}(T^{abc}+T^{bac}-T^{cab})
+\frac{1}{2}(\eta^{ac}T^b-\eta^{ab}T^c).
\end{equation}
The corresponding field equations become
\begin{equation}\label{2.7}
e_{a\lambda}e_{b\mu}\partial_\nu(e\Sigma^{b\lambda\nu})
-e({\Sigma^{b\nu}}_a T_{b\nu\mu}-\frac{1}{4}e_{a\mu}
T_{bcd}\Sigma^{bcd})=\frac{1}{4\kappa}e T_{a\mu},
\end{equation}
where
\begin{equation*}
\frac{\delta L_M}{\delta e^{a\mu}}=e T_{a\mu}.
\end{equation*}

The total Hamiltonian density is \cite{36}
\begin{equation}\label{2.8}
H(e_{ai},\Pi_{ai})=e_{a0}C^a+\alpha_{ik}\Gamma^{ik}+\beta_k\Gamma^k
+\partial_k(e_{a0}\Pi^{ak}),
\end{equation}
where $C^a,~\Gamma^{ik}$ and $\Gamma^k$ are primary constraints,
$\alpha_{ik}$ and $\beta_k$ are the Lagrangian multipliers. In the
constraint $C^a$, the first term is given by a total divergence
\begin{equation}\label{2.9}
C^a=-\partial_i\Pi^{ai}+H^a,\quad \rm{where}~\Pi^{ai}=-4\kappa
e\Sigma^{a0i},
\end{equation}
$C^a$ is the momentum canonically conjugated to $e_{ai}$. The term
$-\partial_i\Pi^{ai}$ is the \textit{energy-momentum density}
\cite{29}. The angular momentum can be defined from the constraint
$\Gamma^{ik}=0$ giving the \textit{angular momentum density}.
\begin{equation}\label{2.12}
2\Pi^{[ik]}=2\kappa
e[-g^{im}g^{kj}{T^0}_{mj}+(g^{im}g^{0k}-g^{km}g^{0i}){T^j}_{mj}].
\end{equation}

The field equations (\ref{2.7}) can also be written as
\begin{equation}\label{2.15}
-\partial_0(\partial_j\Pi^{aj})=-\kappa\partial_j[e
e^{a\mu}(4\Sigma^{bcj}T_{bc\mu}- \delta^j_\mu\Sigma^{bcd}T_{bcd})]
-\partial_j(e{e^a}_\mu T^{j\mu}).
\end{equation}
Integration yields
\begin{equation}\label{2.16}
\frac{d}{dt}[-\int_V d^3x\partial_j\Pi^{aj}]=-\Phi^a_g-\Phi^a_m,
\end{equation}
where
\begin{equation}\label{2.17}
\Phi^a_g=\int_S dS_j\phi^{aj},\quad \Phi^a_m=\int_S dS_j(e{e^a}_\mu
T^{j\mu})
\end{equation}
are the $a$ components of the \textit{gravitational and matter
energy-momentum flux} respectively \cite{37}. $S$ represents the
spatial boundary of the volume $V$. The quantity
\begin{equation}\label{2.19}
\phi^{aj}=\kappa e e^{a\mu}(4\Sigma^{bcj}T_{bc\mu}
-\delta^j_\mu\Sigma^{bcd}T_{bcd})
\end{equation}
is the $a$ component of the gravitational energy-momentum flux
density in $j$ direction. In terms of the gravitational
energy-momentum, Eq.(\ref{2.16}) takes the form
\begin{equation}\label{2.20}
\frac{dP^a}{dt}=-\Phi^a_g-\Phi^a_m.
\end{equation}
Since $P^a=(E,\textbf{P})$, thus \textit{the loss of gravitational
energy} is defined by
\begin{equation}\label{2.21}
\frac{dE}{dt}=-\Phi^{(0)}_g-\Phi^{(0)}_m.
\end{equation}

\section{Stationary Axisymmetric
Einstein-Maxwell Solutions}

It has been the subject of interest over the past decades to study
the behavior of the electromagnetic field in the strong
gravitational field. Electromagnetic fields, in particularly
magnetic fields play a significant role in astrophysics (neutron
stars, white dwarfs and galaxy formation). A comprehensive
relativistic insight of such situations require studying the
Einstein-Maxwell equations. Tupper \cite{38} gave a generalized
metric, which contains five classes of non-null electromagnetic
field plus perfect fluid solutions. Its metric has a symmetry not
inherited by the electromagnetic field. Out of these five classes,
two contain electrovac solutions as special cases and the
remaining three classes must contain fluid. The general form of
the line element describing the stationary axisymmetric solutions
of the Einstein-Maxwell field equations is
$\setcounter{equation}{0}$
\begin{equation}\label{3.1}
ds^2=-dt^2+e^{2K}d\rho^2+(F^2-B^2)d\phi^2+e^{2K}dz^2-2Bdtd\phi,
\end{equation}
where the functions $B=B(\rho,z),~K=K(\rho,z)$ and $F=F(\rho)$
satisfy the following conditions
\begin{eqnarray}\label{3.2}
\dot{B}&=&FW',\quad B'=-F\dot{W},\quad\dot{K}
=-\frac{1}{4}aF(\dot{W}^2-{W'}^2),\nonumber\\
K'&=&-\frac{1}{2}aF\dot{W}W',\quad\ddot{W}+\dot{F}F^{-1}\dot{W}+W''=0.
\end{eqnarray}
Here dot and prime denote differentiation with respect to $\rho$ and
$z$ respectively, $B$ and $F$ have the dimension of length and $K$
is dimensionless, $W$ is an arbitrary function of $\rho$ and $z$,
and $a$ is a constant. The tetrad field corresponding to the line
element (\ref{3.1}) is
\begin{equation}\label{3.3}
{e^a}_\mu(\rho,\phi,z) =\left(\begin{array}{cccc}
1 & 0 & B & 0 \\
0 & e^K\cos\phi & -F\sin\phi & 0 \\
0 & e^K\sin\phi & F\cos\phi & 0 \\
0 & 0 & 0 & e^K \\
\end{array}
\right),
\end{equation}
and its determinant is $e=Fe^{2K}$. The non-vanishing
components of the torsion tensor are
\begin{eqnarray}\label{3.4}
T_{(0)12}&=&-\dot{B},\quad
T_{(0)23}=B',\quad T_{(1)12}=(e^K-\dot{F})\sin\phi,\nonumber\\
T_{(1)13}&=&-K'e^K\cos\phi,\quad
T_{(2)12}=(\dot{F}-e^K)\cos\phi,\quad\nonumber\\
T_{(2)13}&=&-K'e^K\sin\phi,\quad T_{(3)13}=\dot{K}e^K.
\end{eqnarray}
Consequently, the components of the tensor $T_{\lambda\mu\nu}= {e^a}_\lambda
T_{a\mu\nu}$ become
\begin{eqnarray}\label{3.5}
T_{012}&=&-\dot{B},\quad T_{023}=B',\quad T_{113}=-K'e^{2K},\nonumber\\
T_{212}&=&-B\dot{B}+F(\dot{F}-e^K),\quad T_{223}=BB',\quad
T_{313}=\dot{K}e^{2K}.
\end{eqnarray}

\subsection{Energy, Momentum and Angular Momentum}

The non-zero components of energy-momentum density associated with
the metric (\ref{3.1}) are found using Eqs.(\ref{2.9}) and
(\ref{2.6})
\begin{eqnarray}\label{3.6}
-\partial_i\Pi^{(0)i}&=&2\kappa\left(\dot{K}e^K-\ddot{F}-F\ddot{K}-\dot{F}\dot{K}+
\frac{{\dot{B}}^2}{2F}+\frac{B\ddot{B}}{2F}-\frac{B\dot{B}
\dot{F}}{2F^2}+FK''\right.\nonumber\\
&&+\left.\frac{{B'}^2}{2F}+\frac{BB''}{2F}\right),\nonumber\\
-\partial_i\Pi^{(1)i}&=&2\kappa\sin\phi\left(\frac{\ddot{B}}{2}+\dot{B}\dot{K}+
B\ddot{K}-\frac{\dot{B}e^K}{2F}+\frac{B''}{2}-B'K'-BK''\right),\nonumber\\
-\partial_i\Pi^{(2)i}&=&2\kappa\cos\phi\left(-\frac{\ddot{B}}{2}-\dot{B}\dot{K}-
B\ddot{K}+\frac{\dot{B}e^K}{2F}-\frac{B''}{2}+B'K'+BK''\right).
\end{eqnarray}
It is interesting to note that the energy density
$-\partial_i\Pi^{(0)i}$ turn out equivalent to the energy density
obtained in teleparallel theory using the M{\o}ller prescription
\cite{39} and can be re-written as
\begin{equation*}
-\partial_i\Pi^{(0)i}=(-\partial_i{\Pi^{(0)i}})_{GR}+2\kappa(\dot{K}e^K
-\ddot{F}-F\ddot{K}-\dot{F}\dot{K}+FK'').
\end{equation*}
Integration of Eq.(\ref{3.6}) yields energy
\begin{eqnarray}\label{3.7}
P^{(0)}&=&2\kappa\int_Vd^3x\left(\dot{K}e^K-\ddot{F}-F\ddot{K}-\dot{F}\dot{K}+
\frac{{\dot{B}}^2}{2F}+\frac{B\ddot{B}}{2F}-\frac{B\dot{B}\dot{F}}
{2F^2}\right.\nonumber\\
&&+\left.FK''+\frac{{B'}^2}{2F}+\frac{BB''}{2F}\right)
\end{eqnarray}
and momentum vanishes.

The components of angular momentum density $2\Pi^{[ik]}$ are
\begin{eqnarray}\label{3.9}
2\Pi^{[11]}&=&0=2\Pi^{[13]},\quad2\Pi^{[22]}=0=2\Pi^{[33]},\nonumber\\
2\Pi^{[12]}&=&-\frac{2\kappa}{F}(\dot{B}+B\dot{K})=-2\Pi^{[21]},\nonumber\\
2\Pi^{[23]}&=&\frac{2\kappa}{F}(B'-BK')=-2\Pi^{[32]}.
\end{eqnarray}
Consequently, the components of angular momentum become
\begin{eqnarray}\label{3.10}
M^{11}&=&\textmd{constant}=M^{13},\quad M^{22}=\textmd{constant}=M^{33},\nonumber\\
M^{12}&=&-2\kappa\int_Vd^3x\frac{1}{F}(\dot{B}+B\dot{K})=-M^{21},\nonumber\\
M^{23}&=&2\kappa\int_Vd^3x\frac{1}{F}(B'-BK')=-M^{32}.
\end{eqnarray}

\subsection{Energy-Momentum Flux}

Here we discuss gravitational as well as matter energy-momentum
fluxes due to the non-vacuum spacetimes. The components of
gravitational energy flux density $\phi^{(0)j}$ are
\begin{equation}\label{3.11}
\phi^{(0)1}=0=\phi^{(0)2}=\phi^{(0)3}.
\end{equation}
Substituting these values in Eq.(\ref{2.17}) for $a=(0)$, the
gravitational energy flux becomes
\begin{equation}\label{3.12}
\Phi^{(0)}_g=\textmd{constant}.
\end{equation}
The momentum flux $\Phi^{(1)}_g$ for $a=(1)$ gives
\begin{equation}\label{3.13}
\Phi^{(1)}_g=\int_SdS_j\phi^{(1)j},
\end{equation}
where
\begin{eqnarray}\label{3.14}
\phi^{(1)1}&=&2\kappa\cos\phi e^{-K}\left(-\frac{{\dot{B}}^2}{4F}+
F{K'}^2+ \dot{K}e^K-\dot{F}\dot{K}+\frac{{B'}^2}{4F}\right),\nonumber\\
\phi^{(1)2}&=&2\kappa\sin\phi\left(\frac{{\dot{B}}^2}{4F^2}+
\frac{{B'}^2}{4F^2}+{K'}^2\right),\nonumber\\
\phi^{(1)3}&=&2\kappa\cos\phi e^{-K}\left(-\frac{\dot{B}B'}{2F}-
K'e^K+\dot{F}K'\right).
\end{eqnarray}
Replacing these values in Eq.(\ref{3.13}), it follows that
\begin{equation}\label{3.15}
\Phi^{(1)}_g=2\kappa\sin\phi\int_SdS_2\left(\frac{{\dot{B}}^2}
{4F^2}+\frac{{B'}^2}{4F^2}+{K'}^2\right).
\end{equation}
Similarly, we can find $\Phi^{(2)}_g$ using the components of
momentum flux density $\phi^{(2)j}$
\begin{eqnarray}\label{3.16}
\phi^{(2)1}&=&2\kappa\sin\phi e^{-K}\left(-\frac{{\dot{B}}^2}{4F}+
F{K'}^2+ \dot{K}e^K-\dot{F}\dot{K}+\frac{{B'}^2}{4F}\right),\nonumber\\
\phi^{(2)2}&=&-2\kappa\cos\phi\left(\frac{{\dot{B}}^2}{4F^2}+
\frac{{B'}^2}{4F^2}+{K'}^2\right),\nonumber\\
\phi^{(2)3}&=&2\kappa\sin\phi e^{-K}\left(-\frac{\dot{B}B'}{2F}-
K'e^K+\dot{F}K'\right).
\end{eqnarray}
Consequently, we have
\begin{equation}\label{3.17}
\Phi^{(2)}_g=-2\kappa\cos\phi\int_SdS_2
\left(\frac{{\dot{B}}^2}{4F^2}+ \frac{{B'}^2}{4F^2}+{K'}^2\right).
\end{equation}
Finally, the momentum flux
\begin{equation}\label{3.18}
\Phi^{(3)}_g=-\kappa\int_SdS_1\left(\frac{B'\dot{B}e^{-K}}{F}\right)+2\kappa
\int_SdS_3e^{-K}\left(-\frac{{B'}^2}{4F}+FK'+\frac{{\dot{B}}^2}{4F}\right),
\end{equation}
is obtained using the components of momentum flux density
$\phi^{(3)j}$
\begin{eqnarray}\label{3.19}
\phi^{(3)1}&=&-\kappa\left(\frac{B'\dot{B}e^{-K}}{F}\right),\quad
\phi^{(3)2}=0,\nonumber\\
\phi^{(3)3}&=&2\kappa e^{-K}\left(-\frac{{B'}^2}{4F}+FK'
+\frac{{\dot{B}}^2}{4F}\right).
\end{eqnarray}

In order to evaluate matter energy-momentum flux, we have to
calculate matter energy-momentum tensor. Its non-zero components are
\begin{eqnarray}\label{3.20}
T^{00}&=&\frac{e^{-2K}}{8\pi}\left(-\frac{3{\dot{B}}^2}{4F^2}-
\frac{3{B'}^2}{4F^2}+\frac{\ddot{F}}{F}+\ddot{K}+K''+
\frac{7B^2{\dot{B}}^2}{4F^4}+\frac{7B^2{B'}^2}{4F^4}\right.\nonumber\\
&&-\left.\frac{B\dot{B}\dot{F}}{F^3}+\frac{B\ddot{B}}{F^2}+
\frac{BB''}{F^2}-\frac{2B^2\ddot{F}}{F^3}-
\frac{B^2\ddot{K}}{F^2}-\frac{B^2K''}{F^2}\right),\nonumber\\
T^{02}&=&\frac{e^{-2K}}{8\pi}\left(\frac{\dot{B}\dot{F}}{2F^3}-
\frac{\ddot{B}}{2F^2}-\frac{B''}{2F^2}-\frac{B{\dot{B}}^2}{4F^4}
-\frac{B{B'}^2}{4F^4}-\frac{B\ddot{K}}{F^2}-
\frac{BK''}{F^2}\right)=T^{20},\nonumber
\end{eqnarray}
\begin{eqnarray}
T^{11}&=&\frac{e^{-4K}}{8\pi}\left(\frac{{\dot{B}}^2}{4F^2}-
\frac{{B'}^2}{4F^2}+\frac{\dot{F}\dot{K}}{F}\right),\nonumber\\
T^{13}&=&\frac{e^{-4K}}{8\pi}\left(\frac{B'\dot{B}}{2F^2}+
\frac{K'\dot{F}}{F}\right)=T^{31},\nonumber\\
T^{22}&=&\frac{e^{-2K}}{8\pi}\left(\frac{{\dot{B}}^2}{4F^4}+
\frac{{B'}^2}{4F^4}+\frac{\ddot{K}}{F^2}+\frac{K''}{F^2}\right),\nonumber\\
T^{33}&=&\frac{e^{-4K}}{8\pi}\left(-\frac{{\dot{B}}^2}{4F^2}+
\frac{{B'}^2}{4F^2}-\frac{\dot{F}\dot{K}}{F}+\frac{\ddot{F}}{F}\right).
\end{eqnarray}
The energy flux of matter, defined in Eq.(\ref{2.17}), is
\begin{equation*}
\Phi^{(0)}_m=\int_SdS_2e{e^{(0)}}_\mu T^{2\mu}
\end{equation*}
which gives
\begin{equation}\label{3.21}
\Phi^{(0)}_m=\frac{1}{16\pi}\int_SdS_2\left(\frac{\dot{B}\dot{F}}
{F^2}-\frac{\ddot{B}}{F}-\frac{B''}{F}\right).
\end{equation}
The components of momentum flux of matter are
\begin{eqnarray}\label{3.22}
\Phi^{(1)}_m&=&-\frac{1}{8\pi}\sin\phi\int_SdS_2(\frac{{\dot{B}}^2}
{4F^2}+\frac{{B'}^2}{4F^2}+\ddot{K}+K''),\nonumber\\
\Phi^{(2)}_m&=&\frac{1}{8\pi}\cos\phi\int_SdS_2(\frac{{\dot{B}}^2}{4F^2}
+\frac{{B'}^2}{4F^2}+\ddot{K}+K''),\nonumber\\
\Phi^{(3)}_m&=&\frac{1}{8\pi}[\int_SdS_1e^{-K}(\frac{B'\dot{B}}{2F}+
K'\dot{F})+\int_SdS_3e^{-K}(-\frac{{\dot{B}}^2}{4F}+
\frac{{B'}^2}{4F}-\dot{F}\dot{K}+\ddot{F})].\nonumber\\
\end{eqnarray}
Here the non-vanishing values of $\Phi^a_g$ and $\Phi^a_m$
indicate the transfer of gravitational and matter energy-momentum
respectively.

\section{Some Special Cases}

In this section, we evaluate the above results for the two special
cases of stationary axisymmetric Einstein-Maxwell solutions.
\begin{itemize}
\item \textbf{Electromagnetic Generalization of the G\"{o}del
Solution}, which is found by inserting
$B=\frac{m}{n}e^{n\rho},~F=e^{n\rho}$ and $K=0$ in Eq.(\ref{3.1}).
$\setcounter{equation}{0}$
\begin{equation}\label{4.1}
ds^2=-dt^2+d\rho^2+e^{2n\rho}(1-\frac{m^2}{n^2})d\phi^2+dz^2-
\frac{2m}{n}e^{n\rho}dtd\phi,
\end{equation}
\item \textbf{The G\"{o}del Metric}, obtained for
$B=e^{a\rho},~F=\frac{e^{a\rho}}{\sqrt{2}}$ and $K=0$
\begin{equation}\label{4.2}
ds^2=-dt^2+d\rho^2-\frac{1}{2}e^{2a\rho}d\phi^2+dz^2
-2e^{a\rho}dtd\phi,
\end{equation}
where $m,~n$ and $a$ are arbitrary constants with dimension of
$1/L$.
\end{itemize}
The results are summarized in the following table.

\begin{center}
\textbf{Table 1}: Energy and its contents for electromagnetic
generalization of the G\"{o}del solution and the G\"{o}del metric
\end{center}
\begin{center}
\begin{tabular}{|c|c|c|}\hline
\textbf{Quantities} & \textbf{Electromagnetic Generalization} & \textbf{The G\"{o}del Metric} \\
& \textbf{of the G\"{o}del Solution} &  \\\hline $P^{(0)}$ &
$\frac{Le^{n\rho}}{8n}(m^2-2n^2)$ & constant \\\hline $P^{(1)}$ & 0
& 0 \\\hline $P^{(2)}$ & 0 & 0 \\\hline $P^{(3)}$ & constant &
constant \\\hline $\Phi^{(0)}_g$ & constant & constant \\\hline
$\Phi^{(1)}_g$ & $\frac{1}{2}\kappa\rho Lm^2\sin\phi$ & $\kappa
La^2\rho\sin\phi$ \\\hline $\Phi^{(2)}_g$ & $-\frac{1}{2}\kappa\rho
Lm^2\cos\phi$ & $-\kappa La^2\rho\cos\phi$  \\\hline $\Phi^{(3)}_g$
& $\frac{1}{16n}m^2e^{n\rho}-$constant &
$\frac{1}{8\sqrt{2}}ae^{a\rho}-$constant \\\hline $\Phi^{(0)}_m$ &
constant & constant \\\hline $\Phi^{(1)}_m$ & $-\frac{1}{32\pi}\rho
Lm^2\sin\phi$ & $-\frac{1}{16\pi}La^2\rho\sin\phi$ \\\hline
$\Phi^{(2)}_m$ & $\frac{1}{32\pi}\rho Lm^2\cos\phi$ &
$\frac{1}{16\pi}La^2\rho\cos\phi$  \\\hline $\Phi^{(3)}_m$ &
$\frac{1}{16n}(4n^2-m^2)e^{n\rho}+$constant &
$\frac{1}{8\sqrt{2}}ae^{a\rho}+$constant \\\hline
\end{tabular}
\end{center}
It is interesting to mention here that energy for both cases turns
out to be the same as obtained by M{\o}ller's prescription in the
framework of teleparallel gravity \cite{39}. The gravitational and
matter energy fluxes are constant which depict the constant flow of
energy. This indicates that the energy is getting lost at a constant
rate. However, there is an outflow of gravitational momentum along
$\rho$ and $z$ direction, and inward flow along $\phi$ direction.
The components of angular momentum reduce to a constant except
$M^{12}$ which is $-\frac{1}{4}L\rho m$ for the electromagnetic
generalization of the G\"{o}del solution and
$-\frac{1}{2\sqrt{2}}aL\rho$ for the G\"{o}del metric. These
non-zero components of angular momentum show the act of rotation in
both cases.

\section{Cosmic String Spacetimes}

It is generally assumed that the universe in very early stages of
its evolution has gone through a number of phase transitions. The
consequence of this phase transition is the formation of topological
defects, which are associated with spontaneous symmetry breaking.
Cosmic strings are one of the most remarkable defects which are
linear and string like. They have important implications on
cosmology such as large scale structures or galaxy formation. In
this section, energy contents for the non-static ($\Lambda\neq0$)
and static ($\Lambda\neq0$) cosmic strings are evaluated.

Firstly, the energy contents for the \textbf{non-static cosmic}
spacetime are discussed. The non-static line element of the cosmic
string ($\Lambda\neq0$) is \cite{40} $\setcounter{equation}{0}$
\begin{equation}\label{5.1}
ds^2=-dt^2+e^{2\sqrt{\frac{\Lambda}{3}}t}[d\rho^2
+(1-4\mu)^2\rho^2d\phi^2+dz^2],
\end{equation}
where $\mu$ is mass per unit length of the string in geometrized
units ($G=1=c$). The tetrad field is
\begin{equation}\label{5.2}
{e^a}_\mu(t,\rho,\phi)=\left(\begin{array}{cccc}
1 & 0 & 0 & 0 \\
0 & A\cos\phi & -AB\rho\sin\phi & 0 \\
0 & A\sin\phi & AB\rho\cos\phi & 0 \\
0 & 0 & 0 & A \\
\end{array}
\right),
\end{equation}
where $A=e^{\sqrt{\frac{\Lambda}{3}}t}$ and $B=(1-4\mu)$. Its
determinant is $\rho A^3B$. The non-zero components of
$T_{\lambda\mu\nu}$ are
\begin{eqnarray}\label{5.4}
T_{101}=A\dot{A},\quad T_{202}=A\dot{A}B^2\rho^2,\quad
T_{212}=A^2B\rho(B-1),\quad T_{303}=A\dot{A},
\end{eqnarray}
where dot denotes differentiation with respect to $t$.

The energy-momentum density components are
\begin{eqnarray}\label{5.5}
-\partial_i\Pi^{(0)i}&=&0=-\partial_i\Pi^{(3)i},\quad
-\partial_i\Pi^{(1)i}=-16\kappa\mu\sqrt{\frac{\Lambda}{3}}
e^{2\sqrt{\frac{\Lambda}{3}}t}\cos\phi,\nonumber\\
-\partial_i\Pi^{(2)i}&=&-16\kappa\mu\sqrt{\frac{\Lambda}{3}}
e^{2\sqrt{\frac{\Lambda}{3}}t}\sin\phi
\end{eqnarray}
which reflect the symmetry of the cosmic string spacetime. Thus the
energy becomes constant while the momentum vanishes. Also, the
angular momentum turns out to be constant.

\subsection{Energy-Momentum Flux}

The components of gravitational flux density are
\begin{equation}\label{5.8}
\phi^{(0)1}=-8\kappa\mu\sqrt{\frac{\Lambda}{3}}
e^{\sqrt{\frac{\Lambda}{3}}t},\quad\phi^{(0)2}=0=\phi^{(0)3}
\end{equation}
which in turn gives rise to the gravitational energy flux
\begin{equation}\label{5.9}
\Phi^{(0)}_g=-\mu L\sqrt{\frac{\Lambda}{3}}e^{\sqrt{\frac{\Lambda}{3}}t}
+\textmd{constant}.
\end{equation}
Inserting the components of gravitational momentum flux
density $\phi^{(1)i}$,
\begin{equation}\label{5.10}
\phi^{(1)1}=-\frac{2\Lambda}{3}\kappa\rho\cos\phi
e^{2\sqrt{\frac{\Lambda}{3}}t}(1-4\mu),\quad
\phi^{(1)2}=\frac{2\Lambda}{3}\kappa\sin\phi
e^{2\sqrt{\frac{\Lambda}{3}}t},\quad\phi^{(1)3}=0,
\end{equation}
in Eq.(\ref{2.17}) and integrating it over a cylindrical region of
length $L$ and radius $\rho$, we obtain momentum flux
$\Phi^{(1)}_g$
\begin{equation}\label{5.11}
\Phi^{(1)}_g=\frac{2\Lambda}{3}\kappa\sin\phi L\rho
e^{2\sqrt{\frac{\Lambda}{3}}t}+\textmd{constant}.
\end{equation}
Similarly, the momentum flux $\Phi^{(2)}_g$,
\begin{equation}\label{5.12}
\Phi^{(2)}_g=-\frac{2\Lambda}{3}\kappa\cos\phi L\rho
e^{2\sqrt{\frac{\Lambda}{3}}t}+\textmd{constant},
\end{equation}
is found using the components of gravitational momentum flux density
$\phi^{(2)i}$
\begin{equation}\label{5.13}
\phi^{(2)1}=-\frac{2\Lambda}{3}\kappa\rho\sin\phi
e^{2\sqrt{\frac{\Lambda}{3}}t}(1-4\mu),\quad
\phi^{(2)2}=-\frac{2\Lambda}{3}\kappa\cos\phi
e^{2\sqrt{\frac{\Lambda}{3}}t},\quad\phi^{(2)3}=0.
\end{equation}
The momentum flux $\Phi^{(3)}_g$ is
\begin{equation}\label{5.14}
\Phi^{(3)}_g=-\frac{\Lambda}{24}\rho^2
e^{2\sqrt{\frac{\Lambda}{3}}t}(1-4\mu)+\textmd{constant},
\end{equation}
obtained using $\phi^{(3)i}$
\begin{equation}\label{5.15}
\phi^{(3)1}=0=\phi^{(3)2},\quad
\phi^{(3)3}=-\frac{2\Lambda}{3}\kappa\rho
e^{2\sqrt{\frac{\Lambda}{3}}t}(1-4\mu).
\end{equation}

For matter energy-momentum flux, we evaluate the components of the
matter energy-momentum tensor
\begin{eqnarray}\label{5.16}
T^{00}=\frac{\Lambda}{8\pi},\quad T^{11}=-\frac{\Lambda}{8\pi}
e^{-2\sqrt{\frac{\Lambda}{3}}t}=T^{33},\quad
T^{22}=-\frac{\Lambda}{8\pi}
\rho^{-2}B^{-2}e^{-2\sqrt{\frac{\Lambda}{3}}t}.
\end{eqnarray}
Inserting these values in Eq.(\ref{2.17}), it follows that
\begin{eqnarray}\label{5.17}
\Phi^{(0)}_m&=&\textmd{constant},\quad\Phi^{(1)}_m=\frac{\Lambda}{8\pi}\rho
L\sin\phi e^{2\sqrt{\frac{\Lambda}{3}}t}+\textmd{constant},\nonumber\\
\Phi^{(2)}_m&=&-\frac{\Lambda}{8\pi}\rho L\cos\phi
e^{2\sqrt{\frac{\Lambda}{3}}t}+\textmd{constant},\nonumber\\
\Phi^{(3)}_m&=&-\frac{\Lambda}{8}
\rho^2e^{2\sqrt{\frac{\Lambda}{3}}t}(1-4\mu)+\textmd{constant}.
\end{eqnarray}

Now we evaluate the energy contents for the \textbf{static cosmic
string spacetime}. The general form of the static cosmic string
spacetime for $\Lambda\neq 0~(G=1=c)$ in cylindrical polar
coordinate system is \cite{40}
\begin{eqnarray}\label{5.20}
ds^2&=&-\cos^{\frac{4}{3}}(\frac{\sqrt{3\Lambda}}{2}\rho)dt^2+d\rho^2+
\frac{4(1-4\mu)^2}{3\Lambda}\cos^{\frac{4}{3}}(\frac{\sqrt{3\Lambda}}{2}\rho)
\tan^2(\frac{\sqrt{3\Lambda}}{2}\rho)d\phi^2\nonumber\\
&&+\cos^{\frac{4}{3}}(\frac{\sqrt{3\Lambda}}{2}\rho)dz^2.
\end{eqnarray}
The corresponding tetrad field is
\begin{equation}\label{5.21}
{e^a}_\mu(\rho,\phi)=\left(\begin{array}{cccc}
A & 0 & 0 & 0 \\
0 & \cos\phi & -ABC\sin\phi & 0 \\
0 & \sin\phi & ABC\cos\phi & 0 \\
0 & 0 & 0 & A \\
\end{array}
\right),
\end{equation}
where
$A=\cos^{\frac{2}{3}}\alpha,~B=\frac{2(1-4\mu)}{\sqrt{3\Lambda}},~C=\tan\alpha$
and $\alpha=\frac{\sqrt{3\Lambda}}{2}\rho$. The components of
$T_{\lambda\mu\nu}= {e^a}_\lambda T_{a\mu\nu}$ become
\begin{equation}\label{5.23}
T_{001}=A\dot{A},\quad T_{212}=ABC\{B(\dot{A}C+A\dot{C})-1\},\quad
T_{313}=A\dot{A},
\end{equation}
where dot represents derivative with respect to $\rho$.

\subsection{Energy, Momentum and Angular Momentum}

The components of energy-momentum density are
\begin{eqnarray}\label{5.24}
-\partial_i\Pi^{(0)i}&=&-\partial_1[-2\kappa(A-A^2B\dot{C}-2A\dot{A}BC)]\nonumber\\
&=&2\kappa\partial_1[\cos^{\frac{2}{3}}\frac{\sqrt{3\Lambda}\rho}{2}+(1-4\mu)\{\frac{1}{3}
\cos^{-\frac{2}{3}}\frac{\sqrt{3\Lambda}\rho}{2}\nonumber\\
&&-\frac{4}{3}\cos^{\frac{4}{3}}\frac{\sqrt{3\Lambda}\rho}{2}\}],\nonumber\\
-\partial_i\Pi^{(1)i}&=&0=-\partial_i\Pi^{(2)i}=-\partial_i\Pi^{(3)i}.
\end{eqnarray}
The corresponding energy becomes
\begin{eqnarray}\label{5.25}
P^{(0)}&=&\frac{L}{4}[\cos^{\frac{2}{3}}\frac{\sqrt{3\Lambda}\rho}{2}+(1-4\mu)
\{\frac{1}{3}\cos^{-\frac{2}{3}}\frac{\sqrt{3\Lambda}\rho}{2}-\frac{4}{3}
\cos^{\frac{4}{3}}\frac{\sqrt{3\Lambda}\rho}{2}\}],\nonumber\\
&\approx&\mu L+\frac{3L\Lambda^2\rho^2}{24} -\frac{9\mu
L\Lambda^2\rho^2}{16}+\cdots
\end{eqnarray}
and momentum turns out to be constant. Also, the components of
angular momentum become constant.

\subsection{Energy-Momentum Flux}

The gravitational energy flux
\begin{equation}\label{5.26}
\Phi^{(0)}_g=\textmd{constant}
\end{equation}
is found using the components of energy flux density $\phi^{(0)j}$
\begin{equation}\label{5.27}
\phi^{(0)1}=0=\phi^{(0)2}=\phi^{(0)3}.
\end{equation}
The components of the gravitational momentum flux $\Phi^{(i)}_g$
\begin{eqnarray}\label{5.28}
\Phi^{(1)}_g&=&-2\kappa\frac{\Lambda}{3}\sin\phi\int_SdS_2
(\sin^2\frac{\sqrt{3\Lambda}\rho}{2}\cos^{-\frac{2}{3}}
\frac{\sqrt{3\Lambda}\rho}{2})+\textmd{constant},\nonumber\\
\Phi^{(2)}_g&=&2\kappa\frac{\Lambda}{3}\cos\phi\int_SdS_2
(\sin^2\frac{\sqrt{3\Lambda}\rho}{2}\cos^{-\frac{2}{3}}
\frac{\sqrt{3\Lambda}\rho}{2})+\textmd{constant},\nonumber\\
\Phi^{(3)}_g&=&-\frac{1}{4}[\cos^{\frac{2}{3}}
\frac{\sqrt{3\Lambda}\rho}{2}+\frac{1}{3}(1-4\mu)
\cos^{-\frac{2}{3}}\frac{\sqrt{3\Lambda}\rho}{2}
\sin^2\frac{\sqrt{3\Lambda}\rho}{2}]+\textmd{constant},\nonumber\\
\end{eqnarray}
are obtained using the components of the gravitational momentum flux
densities $\phi^{(1)j}$
\begin{eqnarray}\label{5.29}
\phi^{(1)1}&=&2\kappa\sqrt{\frac{\Lambda}{3}}\cos\phi\sin\sqrt{3\Lambda}\rho
\{(1-4\mu)-\cos^{-\frac{2}{3}}\frac{\sqrt{3\Lambda}\rho}{2}\},\nonumber\\
\phi^{(1)2}&=&-2\kappa\frac{\Lambda}{3}\sin\phi\cos^{-\frac{2}{3}}
\frac{\sqrt{3\Lambda}\rho}{2}\sin^2
\frac{\sqrt{3\Lambda}\rho}{2},\nonumber\\
\phi^{(1)3}&=&0,
\end{eqnarray}
$\phi^{(2)j}$,
\begin{eqnarray}\label{5.30}
\phi^{(2)1}&=&2\kappa\sqrt{\frac{\Lambda}{3}}\sin\phi\sin\sqrt{3\Lambda}\rho
\{(1-4\mu)-\cos^{-\frac{2}{3}}\frac{\sqrt{3\Lambda}\rho}{2}\},\nonumber\\
\phi^{(2)2}&=&2\kappa\frac{\Lambda}{3}\cos\phi\cos^{-\frac{2}{3}}
\frac{\sqrt{3\Lambda}\rho}{2}\sin^2
\frac{\sqrt{3\Lambda}\rho}{2},\nonumber\\
\phi^{(2)3}&=&0,
\end{eqnarray}
and $\phi^{(3)j}$
\begin{eqnarray}\label{5.31}
\phi^{(3)1}&=&0=\phi^{(3)2},\nonumber\\
\phi^{(3)3}&=&2\kappa\sqrt{\frac{\Lambda}{3}}\sin\frac{\sqrt{3\Lambda}\rho}{2}
[\cos^{-\frac{1}{3}}\frac{\sqrt{3\Lambda}\rho}{2}-(1-4\mu)\{\frac{1}{3}
\cos^{-\frac{5}{3}}\frac{\sqrt{3\Lambda}\rho}{2}\nonumber\\
&&+\frac{2}{3}\cos^{\frac{1}{3}}\frac{\sqrt{3\Lambda}\rho}{2}\}],
\end{eqnarray}
in Eq.(2.17) for $a=(1),(2)$ and $(3)$ respectively. The non-zero
components of matter energy-momentum tensor are
\begin{eqnarray}\label{5.32}
T^{00}&=&\frac{\Lambda}{8\pi}\cos^{-\frac{4}{3}}
\frac{\sqrt{3\Lambda}\rho}{2},\quad T^{11}=-\frac{\Lambda}{8\pi},\nonumber\\
T^{22}&=&-\frac{3\Lambda^2}{32\pi}(1-4\mu)^{-2}\cos^{\frac{2}{3}}
\frac{\sqrt{3\Lambda}\rho}{2}\sin^{-2}\frac{\sqrt{3\Lambda}\rho}{2},\nonumber\\
T^{33}&=&-\frac{\Lambda}{8\pi}\cos^{-\frac{4}{3}}
\frac{\sqrt{3\Lambda}\rho}{2}.
\end{eqnarray}
Thus we obtain the energy flux of the matter as
\begin{equation}\label{5.33}
\Phi^{(0)}_m=\textmd{constant}
\end{equation}
and the components of momentum flux of matter as
\begin{eqnarray}\label{5.34}
\Phi^{(1)}_m&=&\frac{\Lambda}{8\pi}\sin\phi\int_SdS_2
\cos^{\frac{4}{3}}\frac{\sqrt{3\Lambda}\rho}{2}+\textmd{constant},\nonumber\\
\Phi^{(2)}_m&=&-\frac{\Lambda}{8\pi}\cos\phi\int_SdS_2
\cos^{\frac{4}{3}}\frac{\sqrt{3\Lambda}\rho}{2}+\textmd{constant},\nonumber\\
\Phi^{(3)}_m&=&\frac{1}{4}(1-4\mu)\cos^{\frac{4}{3}}
\frac{\sqrt{3\Lambda}\rho}{2}+\textmd{constant}.
\end{eqnarray}
Substitution of Eqs.(\ref{5.26}) and (\ref{5.33}) in
Eq.(\ref{2.21}) provides the loss of gravitational energy
\begin{equation}\label{5.35}
\frac{dE}{dt}=\textmd{constant}.
\end{equation}

\section{Summary and Discussion}

In this paper, we have evaluated energy, momentum, angular momentum,
gravitational and matter energy-momentum fluxes of stationary
axisymmetric Einstein-Maxwell solutions and cosmic string
spacetimes. The energy expression for stationary axisymmetric
Einstein-Maxwell solutions is well-defined and coincides with the
result of \cite{39} obtained using M{\o}ller's prescription in
teleparallel theory. This also shows consistency with the result of
GR \cite{41} along with some extra terms. However, for some
particular values of $F$ and $K$, it reduces exactly to the energy
expression evaluated in GR \cite{41}, obtained using M{\o}ller's
prescription. Here the momentum components $P^{(i)}$ vanish. All the
components of angular momentum except $M^{12}$ and $M^{23}$ become
constant. The gravitational and matter energy-momentum fluxes are
also investigated. Its non-zero components represent the transfer of
gravitational and matter energy-momentum whose values depend on the
metric coefficients $B,~F$ and $K$. We have also evaluated the above
quantities for the two special cases of stationary axisymmetric
Einstein-Maxwell solutions (electromagnetic generalization of the
G\"{o}del solution and G\"{o}del metric), obtained for specific
values of the metric functions $B,~F$ and $K$.

This procedure has been extended to non-static and static cosmic
strings, for $\Lambda\neq0$. For the static cosmic string, energy
is finite and well-defined whereas momentum turns out to be
constant. For the non-static cosmic string, we have constant
energy. If we take this constant to be zero then the energy
coincides with the result of GR \cite{17}. However, the angular
momentum for both the cases become constant. When
$\Lambda\rightarrow0$, both the metrics (\ref{5.1}) and
(\ref{5.20}) reduce to the static cosmic string ($\Lambda=0$). In
this case, the energy becomes constant and momentum vanishes. When
we choose this constant to be zero, the result agrees with that of
GR \cite{17}. The components of $\Phi^{a}_g$ and $\Phi^{a}_m$ are
well-defined for both cosmic string spacetimes. The flow of energy
is towards the source for non-static cosmic string whereas the
momentum flows outward along radial direction and inwards along
$\phi$ and $z$ direction. For the static cosmic string there is a
constant flow of energy. Thus we can say that the loss of
gravitational energy is constant. For $\Lambda\rightarrow0$, the
gravitational energy-momentum flux becomes constant. However,
matter energy-momentum flux vanishes because the static cosmic
string ($\Lambda=0$) is a vacuum solution of the EFEs. If we set
$\Lambda=0=\mu$ in metrics (\ref{5.1}) and (\ref{5.20}), then the
cosmic string spacetimes reduce to the Minkowski spacetime and
energy-momentum becomes zero as expected due to Minkowski
spacetime.

In the perspective of the above discussion, we can conclude that
the gravitational energy-momentum revealed by the prescription
\cite{29} shows the correspondence with the results of different
energy-momentum complexes, particularly with M{\o}ller's
prescription, both in GR and teleparallel gravity. It is
interesting to mention here that this Hamiltonian approach to
define the conserved quantities yield consistent results in most
of the cases with those already available in the literature. It
would be worthwhile to investigate this problem deeply to get more
comprehension about it which may lead to some indication about its
well-defined and unique solution.

\section*{Appendix}

The non-zero components of the tensor $\Sigma^{abc}$ for

\begin{itemize}
\item \textbf{Stationary Axisymmetric Einstein-Maxwell Solutions}
\begin{eqnarray*}
\Sigma^{001}&=&\frac{e^{-K}}{2}\left(\frac{B\dot{B}}{F^2}e^{-K}+\frac{B^2\dot{K}}{F^2}
e^{-K}+\frac{1}{F}-\frac{\dot{F}}{F}e^{-K}-\dot{K}e^{-K}\right),\\
\Sigma^{003}&=&\frac{e^{-2K}}{2}\left(\frac{BB'}{F^2}-\frac{B^2K'}{F^2}+K'\right),\\
\Sigma^{012}&=&\frac{e^{-2K}}{2F^2}\left(\frac{\dot{B}}{2}+B\dot{K}\right)=-\Sigma^{201},\\
\Sigma^{023}&=&\frac{e^{-2K}}{2F^2}\left(BK'-\frac{B'}{2}\right)=\Sigma^{203},\\
\Sigma^{102}&=&\frac{\dot{B}}{4F^2}e^{-2K},
\quad\Sigma^{113}=-K'e^{-4K},\\
\Sigma^{212}&=&-\frac{\dot{K}}{2F^2}e^{-2K},
\quad\Sigma^{223}=-\frac{K'}{2F^2}e^{-2K},\\
\Sigma^{302}&=&\frac{B'}{4F^2}e^{-2K},\quad\Sigma^{313}=\frac{1}{2F}e^{-3K}(1-\dot{F}e^{-K}),\\
\end{eqnarray*}
\item \textbf{Non-Static Cosmic String}
\begin{eqnarray*}
\Sigma^{001}&=&2\mu\rho^{-1}(1-4\mu)^{-1}e^{-2\sqrt{\frac{\Lambda}{3}}t},\\
\Sigma^{101}&=&\sqrt{\frac{\Lambda}{3}}e^{-2\sqrt{\frac{\Lambda}{3}}t}
=\Sigma^{303},\\
\Sigma^{202}&=&\sqrt{\frac{\Lambda}{3}}\rho^{-2}(1-4\mu)^{-2}
e^{-2\sqrt{\frac{\Lambda}{3}}t},\\
\Sigma^{313}&=&2\mu\rho^{-1}(1-4\mu)^{-1}e^{-4\sqrt{\frac{\Lambda}{3}}t},
\end{eqnarray*}
\item \textbf{Static Cosmic String}
\begin{eqnarray*}
\Sigma^{001}&=&\frac{1}{2\sin\sqrt{3\Lambda}}
\sqrt{\frac{\Lambda}{3}}(1-4\mu)^{-1}\{3+(1-4\mu)(
\cos^{-\frac{4}{3}}\frac{\sqrt{3\Lambda}\rho}{2}\\
&&-4\cos^{\frac{2}{3}}\frac{\sqrt{3\Lambda}\rho}{2})\}=\Sigma^{313},\\
\Sigma^{212}&=&\frac{\sqrt{3}}{4}\Lambda^{\frac{3}{2}}(1-4\mu)^{-2}
\cos^{-\frac{1}{3}}\frac{\sqrt{3\Lambda}\rho}{2}\sin^{-1}\frac{\sqrt{3\Lambda}\rho}{2}.
\end{eqnarray*}
\end{itemize}

\end{document}